\documentclass[11pt]{article}
\usepackage[usenames,dvipsnames]{color}
\usepackage{hyperref}

\begin{document}
\title
{Mass function and particle creation in Schwarzschild-de Sitter spacetime}
\author
{Sourav Bhattacharya{\footnote{souravbhatta[AT]hri.res.in}}
\\
Harish-Chandra Research Institute, Chhatnag Road, Jhunsi, \\
Allahabad-211019, INDIA,\\
and\\
Amitabha Lahiri \footnote{amitabha[AT]bose.res.in}\\
S. N. Bose National Centre for Basic Sciences, \\
Block-JD, Sector III, Salt Lake, Kolkata-700098, INDIA.\\
}

\maketitle
\abstract
We construct a mass or energy function for the non-Nariai class
Schwarzschild-de Sitter black hole spacetime in the region between
the black hole and the cosmological event horizons.
The mass function is local, positive
definite, continuous and increases monotonically with the radial
distance from the black hole event horizon. We derive the
Smarr formula using this mass function, and demonstrate that the
mass function reproduces the two-temperature Schwarzschild-de
Sitter black hole thermodynamics, along with a term corresponding
to the negative pressure due to positive cosmological constant. We
further give a field theoretic derivation of the particle creation
by both the horizons and discuss its connection with the mass
function.

\vskip .5cm

{\bf PACS:} {04.70.Bw, 04.20.Jb}\\
{\bf Keywords:} {Schwarzschild-de Sitter, mass function, Smarr
  formula, particle creation

\vskip 1cm

\section{Introduction}
In general relativity, an important concept is that of a mass
function. It should be regarded as a conserved quantity associated
with the spacetime itself. There are several criteria which
should be obeyed by a quantity if it is to be regarded as a mass
or energy of a given spacetime.  First, it must be defined with
respect to a timelike translational Killing vector field and
second, the mass function must be a positive definite quantity. It
is the latter criterion that makes it a difficult problem to define
a mass function because one cannot construct a satisfactory notion
of conserved gravitational energy-momentum tensor unless one goes
to the asymptotic region~(see e.g.~\cite{Wald:1984rg, weinberg} and
references therein). Only an approximate notion of gravitational
Hamiltonian density can be defined perturbatively and locally but
the positivity of this quantity is far from obvious.

For asymptotically flat spacetimes a gravitational mass can be
defined in several ways. The simplest is the Komar
mass~\cite{Wald:1984rg}. This is proportional to the surface
integral of the derivative of the norm of the timelike Killing
vector field and thus is related directly to geodesic motion. In
general the Komar integral will be positive definite only if the
matter energy-momentum tensor $T_{ab}$ satisfies the strong energy
condition (SEC) : $\left(T_{ab}-\frac{1}{2}T g_{ab}\right) \xi^a
\xi^b \geq 0$, for any timelike $\xi^a$.
 
The second method of defining mass is via the Arnowitt-Deser-Misner
(ADM)~\cite{misner1, misner2, misner3} formalism. In this approach
a gravitational Hamiltonian density is defined with respect to the
background timelike Killing vector field in the asymptotic region
and the integral of this Hamiltonian density is computed. This
integral is interpreted as the gravitational mass.

It is known from the Raychaudhuri equation that geodesics would
converge in a mass distribution only if the latter satisfies the
SEC~\cite{Wald:1984rg, akr, Hawking:1973uf}. Also, it is known that
the SEC usually implies the weak energy condition, i.e. the positivity of
the energy density. Using these two facts a third approach to
define gravitational mass and to prove its positivity was developed
in~\cite{geroch, penrose} for asymptotically flat spacetimes,
assuming every complete null geodesic congruence in the domain of outer
communications admits a pair of conjugate points.

As we mentioned earlier, unlike usual matter fields, the positivity
of the gravitational Hamiltonian density or the gravitational mass
is far from obvious, and hence it requires a formal proof. The
positivity conjecture of the ADM mass was first proved
in~\cite{shon1, shon2}. Soon afterward, a remarkable proof of the
positivity appeared in~\cite{Witten:81} using a spacelike spinor
field on a spacelike non-singular Cauchy surface. This result was
generalized for black holes in asymptotically flat or anti-de
Sitter spacetimes in~\cite{Gibbons:1982jg}. The $\Lambda \leq 0$
spacetimes usually have well defined asymptotic structure or
infinities which are accessible to the geodesic observers. The
references mentioned above consider explicit asymptotic structures
of such spacetimes at spacelike infinities which are uniquely
Minkowskian or anti-de Sitter. In fact the positivity of the ADM
mass for $\Lambda \leq 0$ physically reasonable spacetimes
admitting spin structure and well defined asymptotics is well
understood so far. 

But recent observations suggest that our universe is undergoing a
phase of accelerated expansion~\cite{Riess:1998cb,
  Perlmutter:1998np}, and hence there is a strong possibility that
our universe is endowed with a small but positive cosmological
constant.  We note that, since a positive $\Lambda$ violates SEC,
it repels geodesics (see e.g.~\cite{Rindler:2007zz,Ishak:2008ex,
  Ishak:2008zc,Ishak:2010zh, Schucker:2007}) and hence the first
and the third of the methods mentioned above to define
gravitational mass do not seem to be applicable in this case. Also,
the known exact stationary solutions with $\Lambda>0$ (see
e.g.~\cite{Carter:1968ks}) usually exhibits an outer horizon,
namely the cosmological event horizon. The tiny observed value of
$\Lambda$ (of the order of $10^{-52}{\rm m}^{-2}$) sets the length
scale of the horizon to be ${\cal{O}}
\left(\frac{1}{\sqrt{\Lambda}}\right)$, which is of course very
large, but finite.  If a black hole is present, it will be located
inside the cosmological horizon and the cosmological event horizon
acts in such a spacetime as an outer causal
boundary~\cite{Gibbons:1977mu}, beyond which the timelike Killing
vector field becomes spacelike and communication is not possible
along a future directed causal path thereby ruling out any
meaningful use of asymptotics for an observer located inside the
cosmological horizon.

To the best of our knowledge, one of the earliest construction of
mass in de Sitter black hole spacetimes appeared
in~\cite{Gibbons:1977mu}, where mass function was defined on the
black hole and cosmological horizons using the integral of their
respective surface gravities. The variation of this mass function
gave a Smarr formula. In~\cite{Abbott:82}, metric perturbation was
considered in a region far away from the black hole but inside the
cosmological event horizon where the background spacetime was de
Sitter. A local gravitational energy momentum tensor was
constructed and with respect to the background de Sitter timelike
Killing field the mass of the perturbation was defined. This
perturbative approach has much in common with the usual Hamiltonian
ADM formulation of general relativity.

We note here that although this formalism is not applicable to
spacetimes where the black hole and cosmological horizons are
comparable in size, it is well suited to a universe where black
holes and the cosmological constant are of sizes comparable to
ours.

We shall adopt this approach explicitly in this paper and from now
on call it the Abbott-Deser (AD) formalism. For asymptotically
Schwarzschild-de Sitter spacetimes the mass in this asymptotic
region with respect to the de Sitter background but inside the
cosmological horizon was found to be $M$~\cite{Abbott:82}, i.e. the
mass parameter of the Schwarzschild-de Sitter metric. The spinorial
proof of positivity of ADM mass for asymptotically flat
spacetimes~\cite{Witten:81} was generalized later
in~\cite{Shiromizu:94, Kastor:1996} to show that the mass thus defined
in the sense of~\cite{Abbott:82} with respect to the background de
Sitter spacetime is indeed a positive definite quantity.

What do we wish to do then and with what purpose? The answer is the
following.  Since there exists no preferred asymptotic region in
between the two horizons, there can be no preferred position for an
observer. Accordingly, we shall derive the AD masses in other
regions between the two horizons too, where perturbation scheme is
valid, keeping in mind that firstly the mass must be a continuous
positive definite quantity, and secondly, since positive $\Lambda$
corresponds to a positive energy density, the $\Lambda$ part of
this mass function should increase monotonically with radial
distance from the black hole horizon. We shall not give here a formal proof
for the positive energy theorem in de Sitter spacetimes, but shall
construct a physically reasonable mass function for the Schwarzschild-de
Sitter black hole spacetimes.
 
We outline the basic scheme now. We shall set $c=k_{\rm
B}=G=\hbar=1$ throughout. Let us consider the metric for the
Schwarzschild-de Sitter spacetime written in spherical polar
coordinates,
\begin{eqnarray}
ds^2=-\left(1-\frac{2M}{r}-\frac{\Lambda r^2}{3}\right)dt^2+
\left(1-\frac{2M}{r}-\frac{\Lambda r^2}{3}\right)^{-1}dr^2+
r^2d \theta^2+r^2 \sin^2\theta d\phi^2.
\label{metric}
\end{eqnarray}
For $3M\sqrt{\Lambda}<1$, this spacetime admits three Killing
horizons,
\begin{eqnarray}
r_{\rm{H}}=\frac{2}{\sqrt{\Lambda}}\cos\left[\frac{1}{3}
\cos^{-1}\left(3M\sqrt{\Lambda}\right)+\frac{\pi}{3}\right],~
r_{\rm{C}}=\frac{2}{\sqrt{\Lambda}}\cos\left[\frac{1}{3}
\cos^{-1}\left(3M\sqrt{\Lambda}\right)-\frac{\pi}{3}\right],~r_{\rm{U}}
= -\left(r_{\rm{H}}+r_{\rm{C}}\right).\nonumber \\
\label{s27i}
\end{eqnarray}
$r_{\rm{H}}$ is the black hole event horizon and
$r_{\rm{C}}>r_{\rm{H}}$ is the cosmological horizon, whereas
$r_{\rm{U}}\,,$ being negative, is unphysical.  

As we mentioned earlier, the gravitational mass of the perturbation
over the de Sitter background has been defined and computed earlier
in~\cite{Abbott:82} in a region where $\left(1-\frac{\Lambda
    r^2}{3}\right)\gg \frac{2M}{r}$ and it turned out to be
$M$. Instead, we divide the region between the black hole and the
cosmological event horizon $(r_{\rm{H}}<r <r_{\rm{C}})$ into three
regions of perturbation
\begin{eqnarray}
1\gg\left(\frac{2M}{r}+ \frac{\Lambda r^2}{3}\right)~~({\rm Region
  ~I}), &&\qquad
\left(1-\frac{\Lambda r^2}{3}\right)\gg \frac{2M}{r}~~({\rm Region
  ~II}), \nonumber\\
 \left(1-\frac{2M}{r}\right)\gg \frac{\Lambda r^2}{3}~({\rm Region
   ~III}), && 
\label{pertreg}
\end{eqnarray}
where in the first inequality each term on the right hand side is
much smaller than unity, the term on the right hand side of the
second inequality is much smaller than each of the terms on the
left hand side and similarly for the third inequality. In this sense the
three regions are distinct.  These three regions can respectively
be interpreted as perturbations over background Schwarzschild,
Minkowski and de Sitter spacetimes. For the observed value of
$\Lambda\sim10^{-52}{\rm m}^{-2}$~\cite{Riess:1998cb,
  Perlmutter:1998np}, and $2M$ ranging between the extremes
$10^{4}$m to $10^{14}{\rm m}$, the above regions exist and are
merged smoothly in between. In order to see this explicitly in 
an example, first note that the cosmological horizon is at
$r\sim 10^{26}$m. Let us 
now consider a black hole with $2M\sim 10^4$m. Let us also agree to use the 
symbol $\gg$ to mean that the quantity on the left hand side is more than 
$10^{10}$ times that on the right hand side. Then Region~III extends from
the black hole event horizon till $r\sim 10^{20}$m, the inequality 
breaking down around Planck distance from the horizon. Region~II extends 
from $r\sim 10^{14}$m till about $10^{-12}$m of the cosmological horizon.
Region~I is then from $r\sim 10^{14}$m to $r\sim 10^{21}$m\,. Clearly there 
is considerable overlap between the three regions. A similar 
calculation for $2M\sim 10^{14}$m leads to similar conclusions. 
It is clear that Region~I in Eq.~(\ref{pertreg}) is located in between
the two other regions. Regions~II and III are respectively
repulsion and attraction dominated, whereas Region~I is effectively
flat in the sense that repulsion and attraction nearly balance each
other there. In other words, starting from the side of the black
hole horizon, the consecutive sequence of the above three regions
with increasing radial distance are : III, I, II.

Of course, the above constructions and the AD
formalism are not valid for de Sitter black hole spacetimes for which
the black hole and the cosmological horizons are of comparable sizes. 
But as we have seen, they apply well to the
astrophysical black holes, and for the observed value of the
cosmological constant. Therefore, in order to do physics in the
observed or physical universe, the perturbation scheme described
above should be sufficient.

In particular, we shall compute the gravitational or AD mass
function for each of these regions following~\cite{Abbott:82} in
the next Section.  We shall further see in Section 3 that how the
continuity of the masses in different perturbation regions
leads to the proposal of a new `total' 
mass function. We shall further
present a derivation of the Smarr formula by varying this mass
function and relate this to the spectra of particles created in
this spacetime for a massless quantum scalar field in Section 4.

\section{The mass functions in different
  perturbation regions} 
We start by considering the $\Lambda$-vacuum Einstein equations
\begin{eqnarray}
R_{ab}-\frac{1}{2}R g_{ab} + \Lambda g_{ab}=0,
\label{eins}
\end{eqnarray}
where $R_{ab}$ and $R$ are respectively the Ricci tensor and scalar
computed from the metric.  Let us assume we can find a region in
between the black hole event horizon and the cosmological horizon
where the metric $g_{ab}$ can be decomposed in a background
$g^{(0)}_{ab}$ and a perturbation $\gamma_{ab}$ over it
\begin{eqnarray}
g_{ab}= g^{(0)}_{ab}+ \gamma_{ab},
\label{decompose}
\end{eqnarray}
where $\vert \gamma_{ab}\vert \ll \vert g^{(0)}_{ab}\vert$. The
basic scheme described in~\cite{Abbott:82} can be outlined as
follows : define a local `gravitational energy-momentum tensor'
$T^{\rm{(G)}}_{ab}$ which consists of quadratic and higher order
terms of the $\gamma$'s, whereas the Einstein tensor consists of
$g^{(0)}_{ab}$ and terms linear in $\gamma_{ab}$.  Let
$\nabla^{(0)}_a$ denotes the metric compatible connection on the
background $g^{(0)}_{ab}$. Then the `energy current'
$T^{\rm{(G)}}_{ab}\xi^{(0)a}$ is conserved with respect to the
background, i.e. $\nabla^{(0)}_b \left({T^{\rm{(G)}}}_{a}^{b}
  \xi^{(0)a}\right)\approx0$, where $\xi^{(0)a}$ is the timelike
Killing field corresponding to the background $g^{(0)}_{ab}$,
\begin{eqnarray}
\nabla^{(0)}_a\xi^{(0)}_{b}+\nabla^{(0)}_b\xi^{(0)}_{a}
 \approx 0.
\label{Killing}
\end{eqnarray}
Then one computes the flux of the energy current over a closed
spacelike hypersurface $\Sigma$, orthogonal to $\xi^{(0)}_{a}$. We
shall apply this scheme to compute the gravitational mass of the
perturbations in different regions of Eq.~(\ref{pertreg}).

\subsection{Region~I}
Let us start by considering Region~I of Eq.~(\ref{pertreg}),
i.e. linear or leading perturbation over the Minkowski spacetime
\begin{eqnarray}
g^{(0)}_{tt}&=&-1,~
g^{(0)}_{rr}=1,~
g^{(0)}_{\theta\theta}=r^2,~ g^{(0)}_{\phi\phi}
=r^2\sin^2\theta,\nonumber\\
\gamma_{tt}&=&\left(\frac{2M}{r}+
\frac{\Lambda r^2}{3}\right),
~\gamma_{rr}=\left(\frac{2M}{r}+
\frac{\Lambda r^2}{3}\right),~
\gamma_{\theta\theta}=0=\gamma_{\phi\phi},
\label{pert2}
\end{eqnarray}
and the components of the background Killing vector field,
\begin{eqnarray}
\xi^{(0)\mu}\equiv\left\{1,~0,~0,~0\right\},~~
\xi^{(0)}_\mu\equiv\left\{-1,~
0,~0,~0\right\}.
\label{Killing2}
\end{eqnarray}
The Ricci tensor and scalar reads
\begin{eqnarray}
R_{ac}&=&R^{(0)}_{ac}+
\frac{1}{2}\left[\nabla^{(0)e}\left(\nabla^{(0)}_a
\gamma_{ce}+\nabla^{(0)}_c\gamma_{ae}
-\nabla^{(0)}_e\gamma_{ac}
\right)- \nabla^{(0)}_a\left(\nabla^{(0)f}\gamma_{cf} 
+\nabla^{(0)}_c \gamma-\nabla^{(0)d}\gamma_{cd}
\right)\right]\nonumber\\
&+&{\cal{O}}(\gamma^2)+\dots, \nonumber\\
R&=&R^{(0)} +\left[\nabla^{(0)e}\nabla^{(0)c}\gamma_{ce}
-\nabla^{(0)e}\nabla_{(0)e}\gamma\right]+{\cal{O}}(\gamma^2)
+\dots,
\label{Ricci}
\end{eqnarray}
where the trace is defined with respect to $g^{(0)}_{ab}$, and
$\gamma=\gamma_{ab}g^{(0)ab}$. Also, for the Minkowski background
which is a $\Lambda=0$ vacuum, the Einstein equations become
identities
\begin{eqnarray}
R^{(0)}_{ab}-\frac{1}{2}R^{(0)} 
g^{(0)}_{ab}=0.
\label{minvac}
\end{eqnarray}
Now we use Eq.s~(\ref{Ricci}) and (\ref{minvac}) to expand
Einstein's equations (\ref{eins}). We shift the
${{\cal{O}}{(\gamma^2)}}$ and other higher order terms to the right
hand side of Eq.~(\ref{eins}) which define the gravitational
energy-momentum tensor $\left(8\pi T^{\rm{(G)}}_{ab}-\Lambda
  g_{ab}\right)$ to get
\begin{eqnarray}
\frac{1}{2}\left[\nabla^{(0)d}\nabla^{(0)}_a
\overline{\gamma}_{bd}+\nabla^{(0)d}\nabla^{(0)}_b
\overline{\gamma}_{ad}-\nabla^{(0)d}
\nabla^{(0)}_d\overline{\gamma}
_{ab}-\left(\nabla^{(0)d}\nabla^{(0)c}\overline{\gamma}_{cd}\right)
g^{(0)}_{ab}\right]
=8\pi T^{\rm{(G)}}_{ab}-\Lambda g_{ab},
\label{perteq1}
\end{eqnarray}
where $\overline{\gamma}_{ab}=\gamma_{ab}-\frac{1}{2}\gamma
g^{(0)}_{ab}$.

The gravitational mass $(M_{\rm{G}})$ is defined as the integral of
the `energy current' $\left(8\pi T^{\rm{G}}_{tb}-\Lambda g_{tb}\right)
\xi^{(0)b}$ over a spacelike hypersurface $\Sigma$ orthogonal to
the timelike Killing vector field,
\begin{eqnarray}
M_{\rm{G}}:= \frac{1}{8\pi}\int_{\Sigma} \left(8\pi T^{\rm{(G)}}_{tb}
-\Lambda g_{tb}\right)\xi^{(0)b} d\Sigma^{t},
\label{Mg}
\end{eqnarray}
and `$t$' corresponds to the direction of the timelike Killing
field. Following~\cite{Abbott:82}, we obtain the following
expression for the energy current from Eq.~(\ref{perteq1}) after a
little algebra,
\begin{eqnarray}
\left[T^{\rm{(G)}}_{tb}-\frac{\Lambda}{8\pi}g_{tb}\right]\xi^{(0)b}=
\frac{1}{16\pi}\left[\nabla^{(0)d}
\left(\left(\nabla^{(0)c}H_{tbcd}
\right)\xi^{(0)b}\right)-\nabla^{(0)c}\left(H_{tbcd}
\nabla^{(0)d}\xi^{(0)b}\right)
\right],
\label{current}
\end{eqnarray}
where
\begin{eqnarray}
H_{abcd}=\left(g^{(0)}_{ca}\overline{\gamma}_{bd}-
g^{(0)}_{cd}\overline{\gamma}_{ab}
-g^{(0)}_{ab}\overline{\gamma}_{cd}+
g^{(0)}_{bd}\overline{\gamma}_{ca}\right).
\label{H}
\end{eqnarray}
$H_{abcd}$ is antisymmetric under the interchange of $(a,~d)$ and
$(b,~c)$. Then since in Eq.~(\ref{current}) the indices are fixed
$a=t=b$, i.e. timelike, the indices $(d,~c)$ must be
spacelike. Therefore we can convert the integral in Eq.~(\ref{Mg})
into a surface integral
\begin{eqnarray}
M_{\rm{G}}=
\frac{1}{16\pi} \left[\oint \left(\nabla^{(0)c}H_{tbcd}
\right)\xi^{(0)b}dS^{d}- 
\oint H_{tbcd}\left(\nabla^{(0)d}\xi^{(0)b}\right)dS^{c}\right],
\label{mass}
\end{eqnarray}
where `$dS$' denotes the volume element of a closed 2-surface,
i.e. the boundary of $\Sigma$ of the region of interest
$\Sigma$. Since Schwarzschild-de Sitter spacetime is spherically
symmetric, the closed surface is a 2-sphere. Also,
Eq.~(\ref{pert2}) gives
\begin{eqnarray}
\gamma=\gamma_{ab}g^{(0)ab}=0.
\label{trace}
\end{eqnarray}
We now explicitly evaluate Eq.~(\ref{mass}) using Eq.s
(\ref{pert2}), (\ref{Killing2}) and (\ref{trace}). The covariant
derivative on the background Killing vector field is
$\nabla^{(0)}_{d}\xi^{(0)}_b
=\partial_d\xi^{(0)}_b-\Gamma_{db}^{(0)e}\xi^{(0)}_e$ where we keep
in mind that $d$ and $b$ are summed over as in the equation. Since
$H_{abcd}$ is antisymmetric under the interchange of $a$ and $d$,
and $a$ is timelike, $d$ must be spacelike. Keeping in mind $b=t$,
it is clear that $ \nabla^{(0)}_{d}\xi^{(0)}_b$ is non-vanishing
only when $d=r$. We also have $c=r$ in the second integral of
Eq.~(\ref{mass}). But Eq.s~(\ref{H}), (\ref{pert2}) and
(\ref{trace}) give $H_{ttrr}=0$. Thus the second integrand in
Eq.~(\ref{mass}) is identically vanishing. Now expanding the
covariant derivative in the first integral, keeping in mind
we have to set $b=t$ and $d=r$ after making the expansion,
and using Eq.s~(\ref{H}),~(\ref{pert2}), we find that the only
non-vanishing term is $-g^{(0)ec}\Gamma^{(0)f}_{cr}H_{ttef}$,
where the sum runs over $\theta$ and $\phi$ only, and we finally get
\begin{eqnarray}
M_{\rm{G}}=M+\frac{\Lambda r^3}{6}~.
\label{mass2}
\end{eqnarray}
%
\subsection{Region~II}
Next we consider perturbation over the de Sitter background,
i.e. Region~II : $\left(1- \frac{\Lambda r^2}{3}\right)\gg
\frac{2M}{r}$, in Eq.~(\ref{pertreg}), which was explicitly done
in~\cite{Abbott:82}. The leading perturbation and the background
Killing field are the following
\begin{eqnarray}
g^{(0)}_{tt}&=&-\left(1-\frac{\Lambda r^2}{3}\right),~
g^{(0)}_{rr}=\left(1-\frac{\Lambda r^2}{3}\right)^{-1},~
g^{(0)}_{\theta\theta}=r^2,~ g^{(0)}_{\phi\phi}
=r^2\sin^2\theta,\nonumber\\
\gamma_{tt}&=&\frac{2M}{r},~\gamma_{rr}=
\frac{2M}{r\left(1-\frac{\Lambda r^2}{3}\right)^2},~
\gamma_{\theta\theta}=0=\gamma_{\phi\phi},
\label{pert}
\end{eqnarray}
and
\begin{eqnarray}
\xi^{(0)\mu}\equiv\left\{1,~0,~0,~0\right\},~~
\xi^{(0)}_\mu\equiv\left\{-\left(1-
\frac{\Lambda r^2}{3}\right),~
0,~0,~0\right\}.
\label{Killing}
\end{eqnarray}
The calculation of the mass of the perturbation is essentially the
same as before. The only difference we have to remember is that
unlike the previous case, the de Sitter background now we are
considering now is a $\Lambda$-vacuum
\begin{eqnarray}
R^{(0)}_{ab}-\frac{1}{2}R^{(0)} 
g^{(0)}_{ab}+\Lambda g^{(0)}_{ab}=0.
\label{desitter}
\end{eqnarray}
%
The calculation of the mass of the perturbation, using Eq.s
(\ref{eins}), (\ref{Ricci}) and (\ref{desitter}) and following the
method described above gives
\begin{eqnarray}
M_{\rm{G}}=M.
\label{mass1'}
\end{eqnarray}
Thus the mass of the perturbation with respect to the background de
Sitter spacetime is given by the mass parameter of the
Schwarzschild-de Sitter spacetime.
\subsection{Region~III}
Finally we consider Region~III in Eq.~(\ref{pertreg}),
i.e. perturbation of the $ \Lambda=0$ Schwarzschild background by a
$\Lambda$ term. The background and the perturbation are
\begin{eqnarray}
g^{(0)}_{tt}&=&-\left(1-\frac{2M}{r}\right),~
g^{(0)}_{rr}=\left(1-\frac{2M}{r}\right)^{-1},~
g^{(0)}_{\theta\theta}=r^2,~ g^{(0)}_{\phi\phi}
=r^2\sin^2\theta,\nonumber\\
\gamma_{tt}&=&\frac{\Lambda r^2}{3},~\gamma_{rr}=
\frac{\Lambda r^2}{3\left(1-\frac{2M}
{r}\right)^2},~
\gamma_{\theta\theta}=0=\gamma_{\phi\phi},
\label{pert3}
\end{eqnarray}
and
\begin{eqnarray}
\xi^{(0)\mu}\equiv\left\{1,~0,~0,~0\right\},~
\xi^{(0)}_\mu\equiv\left\{-\left(1-\frac{2M}{r}\right),
~0,~0,~0\right\}.
\label{Killing3}
\end{eqnarray}
We follow the same procedure described after Eq.~(\ref{trace}). The
second integral in Eq.~(\ref{mass}) can be shown to be vanishing as
earlier and evaluating the first integral we get
\begin{eqnarray}
M_{\rm{G}}=
\frac{\Lambda r^3}{6}~,
\label{mass3}
\end{eqnarray}
which is the gravitational mass of the perturbation over the
background Schwarzschild spacetime.
\section{Proposal of a new mass function}
The three AD mass functions $M_{\rm G}$ appearing in
Eq.s~(\ref{mass2}),~(\ref{mass1'}),~(\ref{mass3}) are positive, but
not continuous, since they have been defined with respect to three
different backgrounds. But we have discussed that these three
regions (Eq.~(\ref{pertreg})) are merged smoothly with each
other. This implies that a satisfactory notion of mass function
through these three regions must be continuous as well. In other
words, since the three different background Killing fields are
smoothly merged in the succession of Regions~III,~I and II, any
satisfactory mass function should also share this crucial notion of
continuity. This leads us to propose a new mass function for the
Schwarzschild-de Sitter spacetime, in the following reasonable way,
by taking into account the mass of the background as well. The
phrase `background mass' will be related to the Einstein tensor of
the various gravitational backgrounds described in
Eq.~(\ref{pertreg}), as we shall see at the end of this section.

First we note that for the Minkowski background, the background
curvature is identically vanishing. Therefore we take the
background mass to be zero. For the Schwarzschild background we
define the background mass to be the Komar mass,
\begin{eqnarray}
M_{\rm{B}}=
-\frac{1}{8\pi}\oint \nabla^{(0)}_{a}\xi^{(0)}_{t}dS^a,
\label{komar}
\end{eqnarray}
where the integration is performed over a 2-sphere. For
Schwarzschild background, i.e. in Region~III, $M_{\rm{B}}=M$
anywhere inside that particular perturbation region.

For the de Sitter background we treat the $-\Lambda g^{(0)}_{ab}$
term appearing in the Einstein equations as the energy-momentum
tensor $(8 \pi T_{ab}^{\Lambda})$ corresponding to the cosmological
constant. Thus the corresponding energy current becomes
\begin{eqnarray}
T_{ab}^{\Lambda}\xi^{(0)b}=-\frac{\Lambda}{8 \pi}g^{(0)}_{ab}
\xi^{(0)b}.
\label{lambda}
\end{eqnarray}
But we have from the unperturbed $\Lambda$-vacuum equation
(\ref{desitter}),
\begin{eqnarray}
T_{ab}^{\Lambda}\xi^{(0)b}=-\frac{1}{8 \pi}R^{(0)}_{ab}
\xi^{(0)b}=
\frac{1}{8 \pi}\nabla^{(0)d}\nabla^{(0)}_d\xi^{(0)}_{a},
\label{vacuum}
\end{eqnarray}
using the Killing identity. Thus the gravitational mass
$M_{\rm{B}}$ corresponding to the background de Sitter vacuum is
\begin{eqnarray}
M_{\rm{B}}=\int T_{tb}^{\Lambda}\xi^{(0)b}d \Sigma^t=
\frac{1}{8 \pi}\int \nabla^{(0)d}
\nabla^{(0)}_d\xi_{t}^{(0)}d \Sigma^t.
\label{lambdamass}
\end{eqnarray}
Since $\xi^{(0)a}$ is a timelike coordinate Killing field, the
index $d$ is spacelike above, as can be seen by antisymmetrizing
the covariant derivative on $\xi^{(0)}_t $. So we can convert the
integral in Eq.~(\ref{lambdamass}) into a surface integral over a
2-sphere to get
\begin{eqnarray}
M_{\rm{B}}=\frac{1}{8 \pi}\oint 
\nabla^{(0)}_{d}\xi_{t}^{(0)}dS^{d}
=\frac{\Lambda r^3}{6}~.
\label{lambdamass2}
\end{eqnarray}
Let us now combine the background `gravitational
mass' $M_B$ with the `mass' of the gravitational perturbation $M_G$\, in each 
of the three regions, i.e.
Eq.s~(\ref{mass2}), (\ref{mass1'}), (\ref{mass3}),
with Eq.s~(\ref{komar}), (\ref{lambdamass2}). The result is 
what we may call the {\em total
mass function} $U(r,M)$,
\begin{eqnarray}
U(r,M):=M_{\rm{B}}+M_{\rm{G}}=
M+\frac{\Lambda r^3}{6}~,
\label{total3}
\end{eqnarray}
a formula valid throughout all three perturbation regions. The total 
mass function is positive definite, continuous and monotonically
increases with $r$.

We have seen earlier that the AD mass functions are related to the
gravitational energy density of the perturbation defined over a
given background. We shall now see that the gravitational
masses $M_{\rm B}$ appearing in Eq.s~(\ref{komar}),
(\ref{lambdamass2}) can in fact be related to the `time-time'
component of the Einstein tensor of the associated backgrounds.
To do this, we write $M_{\rm B}$ as
\begin{eqnarray}
M_{\rm{B}}=\frac{1}{8 \pi}\int G^{(0)}_{ta}\xi^{(0)a} d\Sigma^t =
\frac{1}{8 \pi}\int\left[R^{(0)}_{ta}\xi^{(0)a} - \frac12
  R^{(0)}g^{(0)}_{ta}\xi^{(0)a}\right] d\Sigma^t,
\label{massadd''}
\end{eqnarray}
and use the Killing identity to replace the first term on the right hand side 
by $-\nabla^{(0)}_a\nabla^{(0)a}\xi_t$. For Regions~I
and III in Eq.~(\ref{massadd''}), we have $R^{(0)}=0$, whereas for
Region~II we have $R^{(0)}=4\Lambda$. Then for Region~II we may
replace the second term with $2\Lambda
\xi^{(0)}_t=-2\nabla^{(0)}_a\nabla_{(0)}^a\xi_t^{(0)}$ by using Killing's
identity, and eventually arrive at 
Eq.~(\ref{lambdamass2}), whereas
for Region~III we get Eq.~(\ref{komar}).
It is clear that calling the left hand side of
Eq.~(\ref{massadd''}) the background mass is naturally meaningful
due to the appearance of the background Einstein tensor as the
integrand on the right hand side.

As before, putting these all together, the total mass or energy function
$U(r,M)$ appearing in Eq.~(\ref{total3}) for the Schwarzschild-de
Sitter spacetime anywhere in Regions~I,~II,~III, can be written in
a compact and unified form,
\begin{eqnarray}
U(r,M):=M_{\rm{B}}+M_{\rm{G}}=\frac{1}{8 \pi}\int\left[
  R^{(0)}_{ta} - \frac12 R^{(0)}g^{(0)}_{ta}\right]\xi^{(0)a}d\Sigma^t
+M_{\rm{G}}.
\label{massadd}
\end{eqnarray}
We note that $U(r,M)$ is positive definite and monotonically
increasing with $r$, thereby encompassing the satisfaction of weak
energy condition by positive $\Lambda$. 

We have thus seen that mathematically it is justified to refer to the 
function $U(r,M)$ of Eq.~(\ref{total3}) as a local and total mass
or energy function because of the way it is related to the `total'
Einstein tensor (Eq.~(\ref{massadd})), corresponding to the background
(as discussed above) and as well as to the perturbation
(as discussed in the AD formalism in the previous sections). It is interesting to note
also that the $\Lambda$ part of the total mass function $U(r,M)$ is formally similar to
the Tolman-Oppenheimer-Volkoff mass function (see~\cite{Wald:1984rg}
and references therein) for a spherically symmetric general relativistic star.


Let us now also consider
a simple physical example where $U(r,M)$ can naturally be
interpreted as a position dependent mass function associated with
the spacetime. Specifically, we consider gravitational redshift
~\cite{Wald:1984rg, weinberg} 
in Region~I of Eq.~(\ref{pertreg}). For two points $r_1$ and $r_2$ in 
Region~I, we find at leading order that
\begin{eqnarray}
\delta \omega\approx\omega\left(\frac{U(r_1,M)}{r_1} -
  \frac{U(r_2,M)}{r_2}\right), 
\label{massadd'}
\end{eqnarray}
where $\omega$ is the frequency of the photon emitted at $r=r_1$
and $\delta \omega $ is its frequency shift when detected at
$r_2$. We compare this with the result of asymptotically flat 
(i.e. $\Lambda = 0$)
spacetime, in which we get the same formula with $U(r_1,M)=M=U(r_2,M)$, 
and thus it is manifest that in the above equation $U(r,M)$ acts 
as a position dependent mass
function in the Schwarzschild-de Sitter spacetime. We also note that
since $\left(\frac{2M}{r}+\frac{\Lambda r^2}{3}\right)\leq 1$ anywhere in the
region between the two horizons, and 
$\left(1-\frac{2M}{r}-\frac{\Lambda r^2}{3}\right)=1-\frac{2U(r,M)}{r}$,
analogue of Eq.~(\ref{massadd'}) can
be written everywhere in the three perturbation regions,
\begin{eqnarray}
\delta \omega=\omega\left(1-\frac{2U(r_1,M)}{r_1}\right)^{-\frac12}\left[ \left(1-\frac{2U(r_2,M)}{r_2}\right)^{\frac12}-\left(1-\frac{2U(r_1,M)}{r_1}\right)^{\frac12}   \right]. 
\label{massadd''}
\end{eqnarray}
Thus we can see a physical or observational ground of interpreting
$U(r,M)$ to be a position dependent mass or energy function in the
Schwarzschild-de Sitter spacetime.  We instead might have considered the
Komar integral to construct mass function in region~I. But since the Komar mass 
is related to the derivative of the norm of the
timelike Killing vector field (Eq.~(\ref{komar})), it can vanish  
in region~I, as
cosmic repulsion and gravitational attraction nearly balance each
other there.  Consequently, unlike asymptotically flat spacetimes
the Komar mass cannot explain or interpret Eq.~(\ref{massadd'}).
Eq.~(\ref{massadd'}) also gives an example of the physical scenario
where a mass function is required to be constructed in a region
where attraction and repulsion are nearly balancing each other,
thereby providing a further physical justification to the
motivation behind our foregoing calculations. 

We shall see below that our local mass function $U(r,M)$ also
reproduces the thermodynamics for the Schwarzschild-de Sitter black
hole spacetime, but along with a negative pressure term arising due
to positive $\Lambda$.

\section{Smarr formula and particle creation}
Let us compute the variation of this total mass
function $U(r,M)$ (Eq.~(\ref{total3})), subject to the change of the
black hole mass parameter $M$ keeping $\Lambda$ fixed, and will see
that $U(r,M)$ is compatible with the existing idea of
two-temperature de Sitter black hole thermodynamics. For
$\Lambda=0$ stationary black holes, the area theorem and the
constancy of the surface gravity over the event horizon
(see~\cite{Wald:1984rg, Hawking:1973uf} and references therein)
give rise to the idea of black hole 
thermodynamics~\cite{Smarr:1972kt, Hawking:72, Roman, Bekenstein:1973ur, Bardeen}
(see also~\cite{Padmanabhan:2003gd} for a review).

The area of the black hole horizon $(r_{\rm{H}})$ is given by
\begin{eqnarray}
A_{\rm{H}}=4 \pi r_{\rm{H}}^2,
\label{var3}
\end{eqnarray}
which, using Eq.~(\ref{s27i}) we rewrite as
\begin{eqnarray}
M(A_{\rm{H}})=-\frac{4 \Lambda}{3}
\left(\frac{A_{\rm{H}}}{16 \pi}
\right)^{\frac{3}{2}}+\left(\frac{A_{\rm{H}}}{16 \pi}
\right)^{\frac{1}{2}}.
\label{var4}
\end{eqnarray}
Now we write the mass function $U(r,M)$ in terms of two new
variables: the black hole horizon area $A_{\rm{H}}$ and the volume
$V=\frac{4}{3} \pi r^3$ enclosed by a sphere 
of radius $r$ on which we have defined the mass function,
\begin{eqnarray}
U(A_{\rm{H}},V)=-\frac{4 \Lambda}{3}\left
(\frac{A_{\rm{H}}}{16 \pi}
\right)^{\frac{3}{2}}+\left(\frac{A_{\rm{H}}}{16 \pi}
\right)^{\frac{1}{2}}+
\frac{\Lambda V}{8\pi}~,
\label{var5}
\end{eqnarray}
the variation of which gives
\begin{eqnarray}
\delta U(A_{\rm{H}},V)=\left[-\frac{2 \Lambda} {\left(16 
\pi\right)
^{\frac{3}{2}}} \left(A_{\rm{H}}\right)^{\frac{1}{2}}
+\frac{1}{2 \left(16 \pi\right)^{\frac{1}{2}}
\left(A_{\rm{H}}\right)^{\frac{1}{2}}}\right]\delta 
A_{\rm{H}}+\frac{\Lambda}{8\pi}\delta V.
\label{var5}
\end{eqnarray}
The surface gravity $\kappa_{\rm{H}}$ of the black hole horizon is
given by 
\begin{eqnarray}
\kappa_{\rm{H}}=
\left(\frac{M}{{r_{\rm{H}}}^2}
-\frac{\Lambda r_{\rm{H}}}{3}\right),
\label{var6}
\end{eqnarray}
combining which with Eq.s~(\ref{var3}), (\ref{var5}) gives
\begin{eqnarray}
\delta U(A_{\rm{H}},V)=\frac{\kappa_{\rm{H}}}
{8\pi}\delta 
A_{\rm{H}}+\frac{\Lambda}{8\pi}\delta V.
\label{var7}
\end{eqnarray}
A similar calculation with the cosmological horizon yields
\begin{eqnarray}
\delta U(A_{\rm{C}},V)=-\frac{\kappa_{\rm{C}}}
{8\pi}\delta 
A_{\rm{C}}+\frac{\Lambda}{8\pi}\delta V,
\label{var8}
\end{eqnarray}
where $A_{\rm{C}}$ and $\kappa_{\rm{C}}=-\left(\frac{M}{{r_{\rm{C}}}^2}
-\frac{\Lambda r_{\rm{C}}}{3}\right)>0$ are the cosmological
horizon's area and the magnitude of surface gravity respectively. The term
$\frac{\Lambda}{8 \pi}$ appearing in Eq.s~(\ref{var7}) and
(\ref{var8}) can be interpreted as the negative isotropic pressure
due to positive $\Lambda$, and therefore once again justifies the
interpretation of $U(r,M)$ as a physical mass or energy 
function for the spacetime. Eq.s~(\ref{var7}),~(\ref{var8}) thus connect the
variation of our local mass function $U(r,M)$ with the variation of
the horizon parameters.


%
%

In particular, if we now combine 
Eq.s~(\ref{var7}) and (\ref{var8}) for the same volume
$V$\,, we now get a Smarr formula involving horizon parameters only
\begin{eqnarray}
\kappa_{\rm{H}}\delta A_{\rm{H}}+
\kappa_{\rm{C}}\delta A_{\rm{C}}=0\,.
\label{var9}
\end{eqnarray}
This formula was derived earlier in~\cite{Gibbons:1977mu, Kastor:1992nn}, 
and we have rederived it using our local mass function. The Smarr
formula shows that when the area of the black hole horizon increases,
the area of the cosmological horizon decreases and vice versa, which is 
also expected from Eq.s~(\ref{s27i}). 
It is clear from Eq.s~(\ref{var7}),~(\ref{var8})
that, unlike asymptotically flat or anti-de Sitter spacetimes, 
one cannot interpret $\frac{A_{\rm H}}{4}$ alone as the entropy of the
spacetime and instead one defines a `total' 
entropy $S=\frac{1}{4}(A_{\rm H}+A_{\rm C})$~\cite{Kastor:1992nn}. We note
that~\cite{Urano:2009xn, Saida:2009ss, Saida:2011vu} also consider this total entropy
and the variation of $\Lambda$ as well. 

For the non-Nariai class spacetimes that we are considering,
$r_{\rm{H}}\neq
r_{\rm{C}}$, and thus $\kappa_{\rm{H}}\neq \kappa_{\rm{C}}\,$, 
(in particular, $\kappa_{\rm H}>\kappa_{\rm C}$, since $r_{\rm H}<r_{\rm C}$) 
and then Eq.s~(\ref{var7}) and (\ref{var8}) show that unlike
asymptotically flat or anti-de Sitter spacetimes, there is no
unique thermodynamic interpretation in the Schwarzschild-de Sitter
spacetime. However, if we could separate the two 
horizons by a thermally opaque membrane, as considered in
e.g.~\cite{Gibbons:1977mu, Urano:2009xn}, 
more precisely a barrier through which no 
radiation can pass in either direction, we could expect two different
thermal equilibrium states, of temperatures $\frac{\kappa_{\rm
H}}{2\pi}$\,, $\frac{\kappa_{\rm C}}{2\pi}\,,$ corresponding to the
black hole and the cosmological horizons respectively. 
We note that for radiation associated with the black hole
alone, both $\delta M,~\delta A_{\rm H}<0$, whereas for the cosmological
horizon alone, $\delta M >0$ and $\delta A_{\rm C}<0$, in Eq.s~(\ref{var7}),
(\ref{var8}), respectively, which are also indicated by Eq.s~(\ref{s27i}). On the
other hand, the variation of the `total' entropy combined with the
variation of the total mass function $U$ gives
\begin{eqnarray}
\frac{\kappa_{\rm H} \kappa_{\rm C}}{2\pi\left(\kappa_{\rm
      H}-\kappa_{\rm C}\right)} 
\delta\left(\frac{ {A_{\rm H}+A_{\rm C} }} {4}
\right)=-\delta U+\frac{\Lambda}{8\pi}\delta V. 
\label{var10}
\end{eqnarray}
The above variation is also consistent with the 
requirement that $(A_{\rm H}+A_{\rm C})$
increases (decreases) as $M$ decreases (increases) (Eq.s~(\ref{s27i})). 
Eq.~(\ref{var10}) indicates that with
respect to the total entropy one might expect an effective
`equilibrium temperature' $T_{\rm eff}=\frac{\kappa_{\rm H} \kappa_{\rm
C}}{2\pi\left(\kappa_{\rm H}-\kappa_{\rm C}\right)}$. This equilibrium
temperature agrees with that
derived earlier in~\cite{Urano:2009xn} with different mass function
(and sign convention), provided we set $\delta \Lambda=0$
throughout in their derivation.

The interpretation of the $\frac{\Lambda V}{8\pi}$ term as the
pressure-energy can also be found in~\cite{Kastor:2009wy, Dolan:2010ha}, in
the context of anti-de Sitter black holes. The ADM mass parameter was 
interpreted as enthalpy for the AdS black hole by considering variations of 
the cosmological constant, which acts as pressure. 
A similar construction was recently done for de Sitter black 
holes~\cite{Dolan:2013ft}, interprets the term $M+\frac{\Lambda V}{8\pi}$
as a total energy. We note that the constructions in these works were based 
on global Komar integrals defined on the boundaries. Our construction on
the other hand, being based on the AD formalism, produces a local function, 
which the Komar integral does not. 


We shall now derive the different thermal equilibrium
states using Kruskal patches and canonical quantization for an eternal 
Schwarzschild-de Sitter
spacetime corresponding to each of the horizons, and thereby verify
the thermodynamic nature of Eq.s~(\ref{var7}),~(\ref{var8}).
The variation of $U(r,M)$ includes
two things -- the variation at the boundary, and the local variation
(coming from the $\Lambda$ part). Clearly, the variation of energy due to change in
boundary is related to the particle creation effects.

It was shown in~\cite{Hawk} using canonical quantization and
Bogoliubov transformations that a stellar object undergoing
gravitational collapse in an asymptotically flat spacetime to form
a black hole creates Planckian distribution of particles at late
times. For massless quantum fields this distribution can be
measured at the future null infinity, and found to have a
temperature $\frac{\kappa_{\rm H}}{2\pi}$, where $\kappa_{\rm H}$
is the surface gravity of the black hole future horizon. Later this
result was rederived using path integral
quantization~\cite{Hartle:1976tp}. This most remarkable result,
known as Hawking radiation, was further justified by the
renormalization of the quantum energy-momentum tensors
(see~\cite{Birrell, DeWitt:1975ys} and references therein). We
refer our reader to~\cite{Traschen:1999zr, Parker} for
excellent pedagogical reviews on this subject.

For an eternal horizon, there is no scenario for gravitational
collapse and there exists a past horizon, in addition to the future
horizon. It was shown in~\cite{Gibbons:1977mu} using path integrals
that for an eternal Schwarzschild-de Sitter spacetime, the black
hole and the cosmological horizon create thermal particles with
temperatures $\frac{\kappa_{\rm H}}{2\pi}$ and $\frac{\kappa_{\rm
    C}}{2\pi}$ respectively.  In~\cite{Traschen:1999zr}, particle
creation on both the horizons was studied, showing that there can
be non-thermal spectra. In~\cite{Saida:2002mc} particle creation by
a Schwarzschild black hole sitting within a
Friedmann-Robertson-Walker (FRW) universe was studied, in which the
FRW universe could be a global de Sitter space itself. See
also~\cite{Shankaranarayanan:2003ya} for study of particle creation
in Schwarzschild-de Sitter spacetime via complex path analysis.

Let us consider a massless minimally coupled
scalar field $\psi$ moving in the Schwarzschild-de Sitter
spacetime, and ignore any backreaction. Employing the usual
separation of variables, $\psi(t,r,\theta,\phi)=e^{-i\omega
  t}\frac{f_{\omega l m}(r)}{r}Y_{lm}(\theta,\phi)$, the equation
of motion for a single mode becomes
\begin{eqnarray}
-\frac{\partial^2f_{lm}(r,t)}{\partial t^2} +
\frac{\partial^2f_{lm}(r,t)}{\partial r_{*}^2} -
\left(1-\frac{2M}{r}-\frac{\Lambda r^2}{3}\right) 
\left(\frac{l(l+1)}{r^2}+\frac{M}{r^3} - 
\frac{\Lambda}{3}\right)f_{lm}(r,t)=0,
\label{h1}
\end{eqnarray}
where we have abbreviated $f(r,t)=e^{-i\omega t}f(r)$ and
$r_{\star}$ is the tortoise coordinate defined by,
\begin{eqnarray}
r_{\star}=\int \frac{dr}{\left(1-\frac{2M}{r} - \frac{\Lambda
      r^2}{3}\right)}=\frac{1}{2\kappa_{\rm H}}
\ln\left|\frac{r}{r_{\rm H}}-1\right|-
\frac{1}{ 2\kappa_{\rm C}}\ln\left|\frac{r}{r_{\rm C}}-1\right| +
\frac{1}{2\kappa_{\rm U}}\ln\left|\frac{r}{r_{\rm U}}+1\right|,
\label{h2}
\end{eqnarray}
where $\kappa_{\rm U}= \partial_r
\left(1-\frac{2M}{r}-\frac{\Lambda
    r^2}{3}\right)\Big\vert_{r=r_{\rm U}}$. Thus $r_{\star}\to
\mp\infty$ as $r\to r_{\rm H},~r_{\rm C}$ respectively.

For our purpose, we shall first construct suitable coordinate systems
for the Schwarzschild-de Sitter spacetime. There are two coordinate
singularities located at $r_{\rm H}$ and $r_{\rm C}$, therefore we
require two Kruskal-like patches to remove them. We define the
usual outgoing and incoming null coordinates $(u,~v)$ as
\begin{eqnarray}
u=t-r_{\star},~v=t+r_{\star}.
\label{h3}
\end{eqnarray}
By writing the metric (\ref{metric}) in terms of $u$ and $r$, it is
easy to find that $u\to \pm \infty$ as $r\to r_{\rm H},~r_{\rm C}$
respectively along an incoming null geodesic, whereas by writing it
in terms of $v$ and $r$ gives $v\to \mp \infty$ as $ \to r_{\rm
  H},~r_{\rm C}$ respectively along an outgoing null geodesic. In
terms of the null coordinates $(u,~v)$ the metric becomes
\begin{eqnarray}
ds^2=\frac{2M}{r}\left(\frac{r}{r_{\rm H}}-1\right)
\left(\frac{r}{r_{\rm C}}-1\right)\left(\frac{r}{r_{\rm U}} +
1\right)dudv+r^2\left(d\theta^2+\sin^2\theta d\phi^2\right),
\label{h4}
\end{eqnarray}
where $r$ as a function of $(u,~v)$ is understood and can be found
from Eq.~(\ref{h3}). We now define the Kruskal null coordinates for
the black hole event horizon as
\begin{eqnarray}
\overline{u}=-\frac{1}{\kappa_{\rm H}}e^{-\kappa_{\rm H} u},
~\overline{v}=\frac{1}{\kappa_{\rm H}}e^{\kappa_{\rm H} v},
\label{h5}
\end{eqnarray}
so that $\overline{u}\to 0,~-\infty$ as $r\to r_{\rm H},~r_{\rm C}$
respectively, and $\overline{v}\to 0,~+\infty$ as $r\to r_{\rm
  H},~r_{\rm C}$ respectively. The Kruskal null coordinates for the
cosmological event horizon can be defined as
\begin{eqnarray}
\overline{u}'=\frac{1}{\kappa_{\rm C}}e^{\kappa_{\rm C} u},
~\overline{v}'=-\frac{1}{\kappa_{\rm C}}e^{-\kappa_{\rm C} v},
\label{h6}
\end{eqnarray}
so that $\overline{u}'\to +\infty,~0$ as $r\to r_{\rm H},~r_{\rm
  C}$ respectively, and $\overline{v}'\to -\infty,~0$ as $r\to
r_{\rm H},~r_{\rm C}$ respectively. To summarize, the ranges of the
various null coordinates are
\begin{eqnarray}
-\infty <u<\infty,~-\infty <v<\infty,~-\infty <\overline{u}\leq0,
~0\leq\overline{v}<\infty,~0\leq\overline{u}'<\infty,~-
\infty<\overline{v}'\leq 0.
\label{h7e}
\end{eqnarray}
Clearly, there will be both outgoing and incoming mode solutions
for the field equation. Since $(\overline{u},~\overline{u}')$ and
$(\overline{v},~\overline{v}')$ are respectively functions of
$(u,~v)$ only, we have modes in terms of these null coordinates,
\begin{eqnarray}
\psi_{\rm out}&=&a_iu_i+a_i^{\dagger}u_i^{\dagger}=
\overline{a}_i\overline{u}_i+
\overline{a}_i^{\dagger}\overline{u}_i^{\dagger}=
\overline{a}_i'\overline{u}_i'+
\overline{a}_i^{\prime\dagger}\overline{u}_i^{\prime\dagger}\nonumber\\ 
\psi_{\rm in}&=&b_iv_i+b_i^{\dagger}v_i^{\dagger}=
\overline{b}_i\overline{v}_i+
\overline{b}_i^{\dagger}\overline{v}_i^{\dagger}=
\overline{b}_i'\overline{v}_i'+
\overline{b}_i^{\prime\dagger}\overline{v}_i^{\prime\dagger},\nonumber\\
\psi&=&\psi_{\rm in}+\psi_{\rm out},
\label{h7}
\end{eqnarray}
where $(u_i,~v_i)$, $(\overline{u}_i,~\overline{v}_i)$ and
$(\overline{u}_i',~\overline{v}_i')$ are modes corresponding to the
coordinates $(u,~v)$, $(\overline{u},~\overline{v})$ and $(
\overline{u}',~\overline{v}')$ respectively. The index `$i$'
corresponds to all discrete and continuous indices. The complex
quantities $a_i$ etc. are expansion coefficients and as in flat
spacetime, they are interpreted as creation and annihilation
operators associated with respective modes.

The creation and annihilation operators are defined to satisfy the
commutation relations
\begin{eqnarray}
\left[a_i,~a_j^{\dagger}\right]=\delta_{ij}, 
~\left[a_i,~a_j\right]=0=\left[a_i^{\dagger},
~a_j^{\dagger}\right],~\left[b_i,~b_j^{\dagger}\right]=
\delta_{ij},~\left[b_i,~b_j\right]=0=
\left[b_i^{\dagger},~b_j^{\dagger}\right],\nonumber\\
\left[\overline{a}_i,~\overline{a}_j^{\dagger}\right]=
\delta_{ij},~\left[\overline{a}_i,~\overline{a}_j\right]=
0=\left[\overline{a}_i^{\dagger},
~\overline{a}_j^{\dagger}\right],~\left[\overline{b}_i,
~\overline{b}_j^{\dagger}\right]=\delta_{ij},
~\left[\overline{b}_i,~\overline{b}_j\right]=
0=\left[\overline{b}_i^{\dagger},~\overline{b}_j^{\dagger}\right],
\nonumber\\
\quad\left[\overline{a}_i',~\overline{a}_j^{\prime\dagger}\right]=
\delta_{ij},~\left[\overline{a}_i',~\overline{a}_j'\right]=0
=\left[\overline{a}_i^{\prime\dagger},
~\overline{a}_j^{\prime\dagger}\right],
~\left[\overline{b}_i',~\overline{b}_j^{\prime\dagger}\right]=
\delta_{ij},~\left[\overline{b}_i',~\overline{b}_j'\right]=
0=\left[\overline{b}_i^{\prime\dagger},
~\overline{b}_j^{\prime\dagger}\right].
\label{h8}
\end{eqnarray}
The inner product of the modes $(u_i,~v_i)$ are defined as
\begin{eqnarray}
&&(u_i,~u_j)=\frac{i}{2} \int_{\Sigma} \left(u_i^{\dagger}(\nabla_au_j)-u_j(\nabla_au_i^{\dagger})\right)d\Sigma^a=\delta_{ij},~
(v_i,~v_j)=\frac{i}{2} \int_{\Sigma} \left(v_i^{\dagger}(\nabla_av_j)-v_j(\nabla_av_i^{\dagger})\right)d\Sigma^a=\delta_{ij},\nonumber\\
&&(u_i,~u_j^{\dagger})=0=(v_i,~v_j^{\dagger}),
\label{h9}
\end{eqnarray}
where $\Sigma$ is suitable hypersurface and the direction `$a$' is
along its normal. In an asymptotically flat spacetime, one chooses
$\Sigma$ to be the past null infinity. But as we discussed earlier,
in presence of a de Sitter horizon, infinities are not very
meaningful to an observer located within that horizon. So we have
to choose $\Sigma$ differently here.

Let us now define the Bogoliubov transformation coefficients (see e.g.~\cite{DeWitt:1975ys} for details) and consider the outgoing $u$ and $\overline{u}$ modes first,
\begin{eqnarray}
\left(\overline{u}_i,~u_j\right)=\alpha_{ij},~\left(\overline{u}_i,~u_j^{\dagger}\right)=\beta_{ij}.
\label{h11}
\end{eqnarray}
Let us now consider the equality between the first two mode expansions in the first of Eq.s~(\ref{h7}). We use Eq.s~(\ref{h9}), (\ref{h11}) and the commutation relations (\ref{h8}) to get
\begin{eqnarray}
\alpha_{ik}\alpha_{kj}^{\dagger}-\beta_{ik}\beta_{kj}^{\dagger}=\delta_{ij},~\alpha_{ik}\beta_{kj}^{\dagger}-\beta_{kj}^{\dagger}\alpha_{ik}=0.
\label{h13}
\end{eqnarray}
Subject to these relations and the commutations, one can then take the inverse transformations
\begin{eqnarray}
a_i=\alpha_{ij}\overline{a}_j-\beta_{ij}^{\dagger}\overline{a}_j^{\dagger}.
\label{h12}
\end{eqnarray}
If $|0\rangle_{\rm K}$ denotes the vacuum associated with the respective Kruskal modes, then the $(u,~v)$ observer will `see' particles in $|0\rangle_{\rm K}$ in the $i$-th mode as the following,
\begin{eqnarray}
\langle0|_{\rm K}a_i^{\dagger}a_i|0\rangle_{\rm K}=\Sigma_{j}\left|\beta_{ij}\right|^2 ~({\rm no~ sum~ on}~i).
\label{h14}
\end{eqnarray}
Thus, all we have to do now is to determine the Bogoliubov coefficient $\beta_{ij}$.

In order to do that 
we note first that Eq.~(\ref{h1}) admits plane wave solutions infinitesimally close to the horizons in $(u,~v)$
null coordinates,   
\begin{eqnarray}
u(\omega,l,m)\sim \frac{1}{\sqrt{4\pi \omega}}\frac{e^{-i\omega u}}{r}Y_{lm}(\theta,~\phi),\qquad v(\omega,l,m)\sim \frac{1}{\sqrt{4\pi \omega}}\frac{e^{-i\omega v}}{r}Y_{lm}(\theta,~\phi),
\label{h10'}
\end{eqnarray}
whereas near the black hole horizon the mode with respect to $(\overline{u},~\overline{v})$ becomes
\begin{eqnarray}
\overline{u}(\omega,l,m)\sim \frac{1}{\sqrt{4\pi \omega}}\frac{e^{-i\omega \overline{u} }}{r}Y_{lm}(\theta,~\phi),\qquad \overline{v}(\omega,l,m)\sim \frac{1}{\sqrt{4\pi \omega}}\frac{e^{-i\omega \overline{v} }}{r}Y_{lm}(\theta,~\phi),
\label{h10''}
\end{eqnarray}
and near the cosmological horizon the mode with respect to $(\overline{u}',\overline{v}')$ becomes
\begin{eqnarray}
\overline{u}'(\omega,l,m)\sim \frac{1}{\sqrt{4\pi \omega}}\frac{e^{-i\omega \overline{u}' }}{r}Y_{lm}(\theta,~\phi),\qquad \overline{v}'(\omega,l,m)\sim \frac{1}{\sqrt{4\pi \omega}}\frac{e^{-i\omega \overline{v}'}}{r}Y_{lm}(\theta,~\phi),
\label{h10}
\end{eqnarray}
along with their negative frequency counterparts. 

We shall consider a mode outgoing at the future cosmological
horizon and trace in back to the past black hole horizon, where we
shall determine the Bogoliubov coefficients by integrating over the
entire past black hole horizon between this traced back and the
outgoing Kruskal mode in Eq.~(\ref{h10''}).  Consequently, the
surface $\Sigma$ in Eq.~(\ref{h9}) is a closed null hypersurface on
the past black hole horizon. As in the case of asymptotically flat
spacetime~\cite{Hawk, Traschen:1999zr, Parker}, during this
backtracing, there will be some part of the wave which will be
backscattered due to the effective potential barrier in
Eq.~(\ref{h1}) to the future cosmological horizon and hence will be
disconnected from the wave outgoing at the past black hole
horizon. This will be the usual greybody effect associated with the
black hole horizon.

With all the above equipments, our task is now thus to compute
$\beta_{ij}$. We note that on any $r={\rm{constant}}$ hypersurface,
$dt=d(t-r_{\star}(r))=du=e^{\kappa_{\rm H} u}d\overline{u}$, using
Eq.s~(\ref{h5}). There will be a $\partial_{r_{\star}}$ coming from
the normal direction of the hypersurface volume element. But
$\partial_{r_{\star}}e^{-i\omega u}=i\omega e^{-i\omega u}=
-\partial_u e^{-i\omega u}=-e^{-\kappa_{\rm
    H}u}\left(\partial_{\overline{u}}e^{-i\omega
    u(\overline{u})}\right)$, which means for these modes
$dt\partial_{r_{\star}}\equiv
d{\overline{u}}\partial_{\overline{u}}$. Putting these in all
together it is straightforward to calculate
\begin{eqnarray}
\alpha_{\omega,\omega'}=\frac{ik}{4\pi \sqrt{\omega \omega'}
}\int_{(\Sigma,~r=r_{\rm H})}\left[
  \left(\omega'-\frac{\omega}{\kappa_{\rm H}\overline{u}}
  \right)e^{i\omega'\overline{u}}e^{\frac{i\omega}{\kappa_{\rm H}}
    \ln(-\kappa_{\rm H} \overline{u}) }\right]d\overline{u}, 
\label{h15}
\end{eqnarray}
where all the constants including those arising from summation of
the discrete indices and angular integral have been dumped into the
constant $k$. We are yet to choose the limit of the above
integration. The above integration is done on the entire past black
hole horizon, therefore we choose the limit of $\overline{u}$ in
Eq.~(\ref{h15}) to be $-\infty$ to $0$ (Eq.~(\ref{h7e})). With
this, the integral in Eq.~(\ref{h15}) looks exactly the same as in
asymptotically flat spacetime. Analytically continuing this to the
complex plane, and treating $\overline{u}=0$ as a branch cut one
obtains
\begin{eqnarray}
\left|\alpha_{\omega,\omega'}\right|^2=e^{\frac{2\pi
    \omega}{\kappa_{\rm H}}}\left|\beta_{\omega,\omega'}\right|^2. 
\label{h16}
\end{eqnarray}
Then from Eq.~(\ref{h13}) we get 
\begin{eqnarray}
\int d\omega'\Gamma(\omega,\omega')\delta(\omega-\omega')=-k'\int
d\omega'\left(1-e^{\frac{2\pi \omega}{\kappa_{\rm H}}}\right) 
\left|\beta_{\omega,\omega'}\right|^2,
\label{h17}
\end{eqnarray}
where $\Gamma$ stands for possible greybody effect as in the
asymptotically flat spacetimes~\cite{Hawk, Traschen:1999zr} and
$k'$ is some positive constant arising from summation of the
discrete indices and angular integrations.  We note that for
$\kappa_{\rm H}\to 0$, $\beta\to 0$ and there is no particle
creation.  We also note that if instead one computes the Bogoliubov
transformation between two different Kruskal modes, one might have
non-thermal spectra~\cite{Traschen:1999zr}.

Thus the $(u,~v)$ observer will `see' the Kruskal vacuum
corresponding to the black hole horizon is filled with thermal
distribution of outgoing particles
\begin{eqnarray}
I(\omega)\sim \frac{\Gamma(\omega)} { e^{ \frac{2\pi
      \omega}{\kappa_{\rm H} }  }-1}, 
\label{h18}
\end{eqnarray}
with temperature $T_{\rm H}=\frac{\kappa_{\rm H}}{2\pi}$.

A similar analysis can be done for the cosmological event horizon as
the following. We consider ingoing $v$ modes at the future black
hole horizon and incoming $\overline{v}'$ modes at the past
cosmological event horizon. We trace the incoming mode back to the
past cosmological horizon. There will be some part backscatterred
to the black hole and thus generating the greybody effect. We
compute the Bogoliubov transformation coefficient at the past
cosmological event horizon an find thermal spectrum with
temperature $T_{\rm C}=\frac{\kappa_{\rm C}}{2\pi}$. If we set
$\kappa_{\rm C}=0$, there will be no particle creation for the
cosmological horizon.

For fermionic field the commutations are replaced with
anticommutations and Eq.s~(\ref{h13}) is modified with a `$+$' in
place of `$-$'. This will give Fermi-Dirac distribution with the
same respective temperatures.

\section{Discussions}
Let us summarize the results now. We have followed~\cite{Abbott:82} to
construct a mass function in each of the three
different perturbation regions~(Eq.~(\ref{pertreg}))
of the Schwarzschild-de Sitter spacetime. The main motivation
behind this comes from the lack of asymptotic region in between the
black hole and the cosmological horizon. The continuity of
the mass functions in different perturbation regions led us to
define a new, `total' mass function by adding the AD mass with the
mass of the background. We have also shown that such addition is justified since
the mass function thus obtained is related to the 
total Einstein tensor as described earlier.
The resulting final mass function is
positive definite, continuous and its $\Lambda$-part monotonically
increases with the radial distance from the black hole. Thus it
takes care of the weak energy condition satisfied by a positive
$\Lambda$. We have also related our local mass function with the
gravitational redshift effect to give it a natural physical
interpretation.

We note that the perturbation scheme described in
Eq.~(\ref{pertreg}) was the only crucial ingredient for our
calculations. Such scheme is clearly not valid for Schwarzschild-de
Sitter spacetimes with comparable sizes of horizons, but as we have
argued earlier, to do physics in the universe we live in, such
construction is reasonable and should be sufficient.

The most useful feature of this mass function is manifest in
particular when we consider Region~I in Eq.~(\ref{pertreg}) as the
following. Since the norm of the timelike Killing vector field
vanishes at the two horizons, it reaches a maximum in between,
where a geodesic in nearly undeflected.  Clearly, this region
corresponds to Region~I. Hence if there is a geodesic observer,
he/she will `feel' that there is no mass within of the spacetime at
all. This of course cannot be acceptable.

We have rederived the two-temperature thermodynamic relations by
varying this mass function for the Schwarzschild-de Sitter
spacetime.  Apart from the surface gravity terms, we have obtained
a term due to negative isotropic pressure exerted by a positive
$\Lambda$. This once again justifies the physical validity of our
mass function.
  
Finally we have computed particle creation in this spacetime for
both the horizons using canonical quantization. Thus the
thermodynamic nature of Eq.s~(\ref{var7}), (\ref{var8}) are
verified. The $\frac{\Lambda \delta V}{8\pi}$ term should be
regarded as change in the (negative) pressure-energy measured by an observer due to the
infinitesimal displacement of the observer from his/her initial
position, which is purely local, in contrast to the asymptotically
flat spacetimes. Thus the mass function we constructed takes care
of the local variation of energy of the ambient de Sitter spacetime as
well, which in fact can be comparable with the energy of the
created particles.

\section*{Acknowledgment}
The authors thank anonymous referees for critically reading the 
manuscript. SB thanks Amit Ghosh for useful comments and discussions.


\bigskip

\begin{thebibliography}{99} 

\bibitem{Wald:1984rg}
  R.~M.~Wald,
  ``General Relativity,''
{\it  Chicago Univ. Pr. ( 1984)}.

\bibitem{weinberg}
  S.~Weinberg,
  ``Gravitation and Cosmology,''
{\it  John Wiley and Sons, New York (1972)}.



\bibitem{misner1}
R.~ Arnowitt, S.~Deser and C.~W.~Misner, 
Phys. \ Rev. \ {\bf 117}, 1595 (1960). 



\bibitem{misner2}
R.~ Arnowitt, S.~Deser and C.~W.~Misner, 
Phys. \ Rev. \ {\bf 118}, 1100 (1960). 


\bibitem{misner3}
R.~ Arnowitt, S.~Deser and C.~W.~Misner, 
Phys. \ Rev. \ {\bf 122}, 997 (1961). 

\bibitem{akr}
A.~K.~Raychaudhuri, 
Phys. \ Rev. \ {\bf 98}, 1123 (1955). 


\bibitem{Hawking:1973uf}
  S.~W.~Hawking and G.~F.~R.~Ellis,
  ``The Large scale structure of spacetime,''
{\it  Cambridge Univ. Pr. (1973)}.


\bibitem{geroch}
R.~Geroch and G.~T.~Horowitz,
Ann. \ Phys. \  {\bf 117}, 1 (1979).



\bibitem{penrose}
R.~Penrose, R.~D.~Sorkin, E.~Woolgar,
arXiv:gr-qc/9301015.


\bibitem{shon1}
R.~Schon and S.~T.~Yau, 
Commun.\ Math. \ Phys. \ {\bf 65}, 45 (1979).


\bibitem{shon2}
R.~Schon and S.~T.~Yau, 
Commun.\ Math. \ Phys. \ {\bf 79}, 231 (1981).


\bibitem{Witten:81}
  E.~Witten,
 Commun.\ Math.\ Phys.\ {\bf 80}, 381 (1981).

\bibitem{Gibbons:1982jg} 
  G.~W.~Gibbons, S.~W.~Hawking, G.~T.~Horowitz and M.~J.~Perry,
  Commun.\ Math.\ Phys.\  {\bf 88}, 295 (1983).

\bibitem{Riess:1998cb}
  A.~G.~Riess {\it et al.}  [Supernova Search Team Collaboration],
  Astron.\ J.\  {\bf 116}, 1009 (1998).


\bibitem{Perlmutter:1998np}
  S.~Perlmutter {\it et al.}  [Supernova Cosmology Project Collaboration],
  Astrophys.\ J.\  {\bf 517}, 565 (1999).


\bibitem{Rindler:2007zz}
  W.~Rindler and M.~Ishak,
  Phys.\ Rev.\  D {\bf 76}, 043006 (2007).



\bibitem{Ishak:2008ex}
M.~Ishak,
Phys.\ Rev.\  D {\bf 78}, 103006 (2008).


\bibitem{Ishak:2008zc}
  M.~Ishak, W.~Rindler and J.~Dossett,
  Mon.\ Not.\ Roy.\ Astron.\ Soc.\  {\bf 403}, 2152 (2010).


\bibitem{Ishak:2010zh}
  M.~Ishak and W.~Rindler,
  Gen.\ Rel.\ Grav.\  {\bf 42}, 2247 (2010).


\bibitem{Schucker:2007}
  T.~Schucker, 
Gen.\ Rel. \ Grav.\ {\bf 41}, 67 (2009).

\bibitem{Carter:1968ks}
  B.~Carter,
  Commun.\ Math.\ Phys.\  {\bf 10}, 280 (1968).


\bibitem{Gibbons:1977mu}
  G.~W.~Gibbons and S.~W.~Hawking,
  Phys.\ Rev.\  D {\bf 15}, 2738 (1977).



\bibitem{Abbott:82}
  L.~F.~Abbott and S.~Deser,
  Nucl.\ Phys.\ B \ {\bf 195}, 76 (1982).



\bibitem{Shiromizu:94}
T.~Shiromizu, 
Phys. \ Rev. \ D {\bf 49}, 5026 (1994).



\bibitem{Kastor:1996}
D.~Kastor and J.~Traschen,
Class. \ Quant. \ Grav. {\bf 13}, 2753 (1996).


\bibitem{Smarr:1972kt}
  L.~Smarr,
  Phys.\ Rev.\ Lett.\  {\bf 30}, 71 (1973)
  [Erratum-ibid.\  {\bf 30}, 521 (1973)].

\bibitem{Hawking:72}
 S.~W.~Hawking, 
 Commun.\ Math.\ Phys. \ {\bf 25}, 152 (1972).

\bibitem{Roman}
T.~A.~Roman,
Gen.\ Rel.\ Grav.\ {\bf 20}, 359 (1988).

\bibitem{Bekenstein:1973ur}
  J.~D.~Bekenstein,
  Phys.\ Rev.\  D {\bf 7}, 2333 (1973).


\bibitem{Bardeen}
J.~M.~Bardeen, B.~Carter and S.~W.~Hawking,
Commun.\ Math.\ Phys.\ {\bf 31}, 161 (1973). 



\bibitem{Padmanabhan:2003gd}
  T.~Padmanabhan,
  Phys.\ Rept.\  {\bf 406}, 49 (2005).

\bibitem{Kastor:1992nn} 
  D.~Kastor and J.~H.~Traschen,
  Phys.\ Rev.\ D {\bf 47}, 5370 (1993).



\bibitem{Urano:2009xn} 
  M.~Urano, A.~Tomimatsu and H.~Saida,
  Class.\ Quant.\ Grav.\  {\bf 26}, 105010 (2009).


\bibitem{Saida:2009ss} 
  H.~Saida,
  Prog.\ Theor.\ Phys.\  {\bf 122}, 1515 (2010).


\bibitem{Saida:2011vu} 
  H.~Saida,
  arXiv:1109.0801 [gr-qc].


\bibitem{Kastor:2009wy} 
  D.~Kastor, S.~Ray and J.~Traschen,
  Class.\ Quant.\ Grav.\  {\bf 26}, 195011 (2009).


\bibitem{Dolan:2010ha} 
  B.~P.~Dolan,
  Class.\ Quant.\ Grav.\  {\bf 28}, 125020 (2011).

\bibitem{Dolan:2013ft} 
  B.~P.~Dolan, D.~Kastor, D.~Kubiznak, R.~B.~Mann and J.~Traschen,
  Phys.\ Rev.\ D {\bf 87}, 104017 (2013).



\bibitem{Hawk}
S.~W.~Hawking,
 Commun.\ Math.\ Phys. \ {\bf 43}, 199 (1975).


\bibitem{Hartle:1976tp}
  J.~B.~Hartle and S.~W.~Hawking,
  Phys.\ Rev.\  D {\bf 13}, 2188 (1976).

\bibitem{Birrell}
  N.~D.~Birrell and P.~C.~W.~Davies,
  ``Quantum fields in curved space,''
{\it  Cambridge Univ. Pr. (1982).}


\bibitem{DeWitt:1975ys} 
  B.~S.~DeWitt,
  Phys.\ Rept.\  {\bf 19}, 295 (1975).

\bibitem{Traschen:1999zr} 
  J.~H.~Traschen,
  gr-qc/0010055.


\bibitem{Parker}
  L.~Parker and D.~Toms,
  ``Quantum Field Theory in Curved Spacetime,''
{\it  Cambridge Univ. Pr. (2009)}.


\bibitem{Saida:2002mc} 
  H.~Saida,
  Class.\ Quant.\ Grav.\  {\bf 19}, 3179 (2002).



\bibitem{Shankaranarayanan:2003ya} 
  S.~Shankaranarayanan,
  Phys.\ Rev.\ D {\bf 67}, 084026 (2003).









\end{thebibliography}
\end{document}